\documentclass[conference]{IEEEtran}
\IEEEoverridecommandlockouts
\usepackage{cite}
\usepackage{amsmath,amssymb,amsfonts}
\usepackage{algorithmic}
\usepackage{multirow}
\usepackage{graphicx}
\usepackage{textcomp}
\usepackage[section]{placeins}
\usepackage{float}
\usepackage{xcolor}
\usepackage{url}

\def\BibTeX{{\rm B\kern-.05em{\sc i\kern-.025em b}\kern-.08em
    T\kern-.1667em\lower.7ex\hbox{E}\kern-.125emX}}
\begin{document}

\title{Low Error-Rate Approximate Multiplier Design for DNNs with Hardware-Driven Co-Optimization\\
\thanks{This work is supported by the National Natural Science Foundation of China under grants 61971143 and 62174035. }
}

\author{\IEEEauthorblockN{Yao Lu, Jide Zhang, Su Zheng, Zhen Li,  Lingli Wang}
\IEEEauthorblockA{State Key Laboratory of ASIC \& System, Fudan University, Shanghai, China \\
Emails: \{yaolu20, szheng19, lizhen19, llwang\}@fudan.edu.cn, zhangjd21@m.fudan.edu.cn}
}

\maketitle

\begin{abstract}
In this paper, two approximate 3×3 multipliers are proposed and the synthesis results of the ASAP-7nm process library justify that they can reduce the area by 31.38\% and 36.17\%, and the power consumption by 36.73\% and 35.66\% compared with the exact multiplier, respectively. They can be aggregated with a 2×2 multiplier to produce an 8×8 multiplier with low error-rate based on the distribution of DNN weights. We propose a hardware-driven software co-optimization method to improve the DNN accuracy by retraining. Based on the proposed two approximate 3-bit multipliers, three approximate 8-bit multipliers with low error-rate are designed for DNNs. Compared with the exact 8-bit unsigned multiplier, our design can achieve a significant advantage over other approximate multipliers on the public dataset.
\end{abstract}


\section{Introduction}
Approximate computing has gradually become an energy-saving solution for digital systems \cite{b1}. In some error-tolerant applications that do not require high precision, such as image processing, communication system, deep neural networks (DNNs), etc., approximate arithmetic circuits can usually bring the hardware cost advantages such as the area, power, and critical path delay. In DNN applications, the accuracy loss caused by approximation can be negligible by the careful co-optimization of approximate multiplier architectures with the DNN retraining software platform.

References \cite{b2} has shown that approximate multipliers can effectively reduce the area and power consumption with small precision loss.  It is a common way to approximate multiplication by algebraic methods. A logarithm-based multiplication system is introduced in \cite{b3}, which approximates the logarithm by the Mitchell Algorithm through a linear function. An iterative logarithmic multiplier is introduced in \cite{b4}, which continuously approaches the exact value by controlling the number of decimal iterations. A further derived truncated error correction \cite{b5} is applied to the truncated iterative multiplier in \cite{b6}, which provides greater flexibility than the static approximate multiplication, with the improvement of latency, precision, and hardware cost. 

Moreover, the research on approximate multiplication is no longer limited to arithmetic algorithms but hardware architectures. Reference \cite{b7} describes in detail a low-power, high-performance approximate multiplier, SiEi, which improves accuracy by compensating some of the errors. A multiplier RoBA in \cite{b8} has the characteristics of high speed and energy saving. It rounds the operand to the nearest exponent of two. This method saves the intensive part of the multiplication operations at the cost of a high error rate. The error-tolerant multiplier (ETM) of \cite{b9} is based on the truncation of a multiplier into an accurate multiplication part for most-significant-bits (MSBs) and a non-multiplication part for least-significant-bits (LSBs). The method of modifying the low-bit width multiplier based on the Karnaugh map (K-map) proposed in \cite{b10} has been proven to effectively reduce the area and critical path delay. This method is also used in \cite{b11}\cite{b12} for the approximate 4-2 compressor, full adder, half adder, and then applied in the reduction process of the Wallace multiplier tree. 

In \cite{b13}\cite{b14}, there are different error metrics for approximate multipliers: error distance (\textbf{ED}), mean error distance (\textbf{MED}), normalized mean error distance (\textbf{NMED}), mean relative error distance (\textbf{MRED}) and error rate (\textbf{ER}). These metrics are useful to evaluate multipliers but not necessarily suitable for various applications. Hence, DNN accuracy loss (\textbf{DAL}) is adopted to reflect the accuracy of DNN caused by approximate multipliers.

Through the aggregation of low bit-width multipliers, the Wallace multiplier can effectively reduce the layer number of adders to shorten the critical path. Reference \cite{b10} introduces a 2×2 approximate multiplier, which has significant improvement in both area and power consumption but leads to a high accuracy loss after aggregating into a large multiplier. Hence two 3×3 approximate multiplier architectures are proposed, which can be used for the partial product generation for aggregating into large multipliers according to the distribution of DNN weights. In DNN accelerators, 8-bit is a common configuration. The results of the 8-bit quantization experiment in \cite{b15} are convictive and Eyeriss-v2 in \cite{b16} also uses 8-bit quantization configuration. Three 8×8 unsigned multiplier architectures are then proposed based on 3-bit multipliers and evaluated by the extended DNN platform based on \cite{b17}. 

With the proposed approximate multipliers, the evaluation results show that the average \textbf{DAL} is about 0.4\% and 12\% for LeNet on MNIST dataset \cite{b18} and CIFAR10  dataset \cite{b19}, respectively. Our platform then retrains the DNN by regularization to improve DNN accuracy, which can help reduce the average accuracy loss to 0.2\% and 9\%. In addition, VGG16, AlexNet and ResNet-19 are adopted on CIFAR10 to evaluate the \textbf{DAL} for larger DNNs. Our main contributions are as follows: 

\begin{itemize}
\item Two approximate 3×3 multipliers with small area and low critical path delay are proposed for multiplier aggregation.
\item Three approximate 8×8 multipliers with low ER and NMED are designed by aggregating two approximate 3×3 multipliers with one 2×2 multiplier according to the distribution of DNN weights.
\item The DNN platform is extended to evaluate the \textbf{DAL} caused by approximate 8×8 multipliers. Moreover, DNN accuracy can be improved by retraining.
\end{itemize}

The rest of the paper is organized as follows. Section \ref{sec2} discusses the approximation method of the 3×3 multiplier architecture and the aggregation method of an 8-bit unsigned multiplier. In Section \ref{sec3}, the proposed multiplier is compared with the exact multiplier and existing approximate multipliers in terms of area, power, delay, \textbf{ER}, \textbf{MED}, \textbf{NMED} and \textbf{MRED}. In Section \ref{sec4}, DNN retraining is introduced to reduce the DNN accuracy loss and the approximate multiplier is evaluated by the DNN platform. The conclusion and our future work are discussed in Section \ref{sec5}.

\section{Proposed Architectures of Approximate Multipliers}\label{sec2}
\subsection{Approximate 3×3 Multipliers}
A 3×3 multiplier has 6 inputs $\alpha_{2,1,0}$, $\beta_{2,1,0}$ and 6 outputs $O_{5,4,3,2,1,0}$  with complicated logic functionality and large area and delay cost. For 6 inputs, there are 64 (2\textsuperscript{6}) truth-table values. Only when the value is more than 31 ($O_{5}$=1), the hardware architecture needs to use the sixth output $O_{5}$=1. By modifying its K-map, it can simplify the logic  functionality with an error rate constraint. 
As shown in TABLE~\ref{tab1}, there are total 6 values that are larger than 31 in the truth table. Therefore, modifying these six cases to satisfy $O_{5}$=0 can cut down the output bits to 5, which can save the area. TABLE~\ref{tab2} lists the modified truth table to have the first design MUL$_{3\times3}$\_1.
In the table, \textbf{Value} and \textbf{Value'} represent exact values and approximate values, respectively.

\begin{equation}
ED=\left\vert Value'-Value \right\vert\label{eq1}
\end{equation}
\begin{equation}
MED=\frac{\sum_{i=1}^{2^n} ED}{2^n}\label{eq2}
\end{equation}
\begin{equation}
ER=\frac{m}{2^n}\label{eq3}
\end{equation}

The expressions for \textbf{ED}, \textbf{MED}, and \textbf{ER} are shown in \eqref{eq1}-\eqref{eq3}, where $2^n$  is the total number of results produced by different inputs, $n$ is the bit width of a multiplier, and $m$ is the number of approximate values. 
From \eqref{eq2}\eqref{eq3}, the \textbf{ER} and \textbf{MED} of MUL$_{3\times3}$\_1 are 9.375\% ($\tfrac{6}{64}$) and 1.125, respectively.

\begin{table}[htbp]
\caption{Truth Table for 3×3 Exact Multiplier (Value\textgreater31)}
\begin{center}
\renewcommand{\arraystretch}{1.2}
\begin{tabular}{|c|c|c|c|c|c|c|c|c|}
\hline
\textbf{$\alpha_{2,1,0}$} & \textbf{$\beta_{2,1,0}$}&\textit{\textbf{Value}}&\textbf{$O_5$}&\textbf{$O_4$}&\textbf{$O_3$}&\textbf{$O_2$}&\textbf{$O_1$}&\textbf{$O_0$}\\
\hline
101&111&35&1&0&0&0&1&1\\
\hline
110&110&36&1&0&0&1&0&0\\
\hline
110&111&42&1&0&1&0&1&0\\
\hline
111&101&35&1&0&0&0&1&1\\
\hline
111&110&42&1&0&1&0&1&0\\
\hline
111&111&49&1&1&0&0&0&1\\
\hline
\end{tabular}
\label{tab1}
\end{center}
\end{table}
\begin{table}[htbp]
\caption{Approximate Truth Table for MUL$_{3\times3}$\_1 (Value\textgreater31)}
\begin{center}
\resizebox{\columnwidth}{!}{
\renewcommand{\arraystretch}{1.5}
\begin{tabular}{|c|c|c|c|c|c|c|c|c|c|c|}
\hline
\textbf{$\alpha_{2,1,0}$} & \textbf{$\beta_{2,1,0}$}&\textit{\textbf{Value}}&\textbf{$O_5$}&\textbf{$O_4$}&\textbf{$O_3$}&\textbf{$O_2$}&\textbf{$O_1$}&\textbf{$O_0$}&\textit{\textbf{Value'}}&\textit{\textbf{ED}}\\
\hline
101&111&35&0&1&1&0&1&1&27&8\\
\hline
110&110&36&0&1&1&0&0&0&24&12\\
\hline
110&111&42&0&1&1&1&1&0&30&12\\
\hline
111&101&35&0&1&1&0&1&1&27&8\\
\hline
111&110&42&0&1&1&1&1&0&30&12\\
\hline
111&111&49&0&1&1&1&0&1&29&20\\
\hline
\end{tabular}
}
\label{tab2}
\end{center}
\end{table}

From TABLE \ref{tab2}, logic functionality expressions \eqref{eq4}-\eqref{eq9} can be derived through the software\cite{b20}. 

\begin{equation}
O_0=\alpha_0\beta_0\label{eq4}
\end{equation}
\begin{equation}
O_1=\Bar{\alpha_1}\alpha_0\beta_1+\alpha_1\Bar{\alpha_0}\beta_1+\alpha_1\Bar{\beta_1}\beta_0+\alpha_0\beta_1\Bar{\beta_0}\label{eq5}
\end{equation}
\begin{equation}
\begin{split}
O_2=&\Bar{\alpha_2}\alpha_1\Bar{\alpha_0}\beta_1+\alpha_1\Bar{\beta_2}\beta_1\Bar{\beta_0}+\Bar{\alpha_2}\Bar{\alpha_1}\alpha_0\beta_2\\
&+\Bar{\alpha_2}\alpha_0\beta_2\Bar{\beta_1}+\alpha_1\beta_2\beta_1\beta_0+\alpha_2\Bar{\alpha_1}\Bar{\alpha_0}\beta_0\\
&+\alpha_2\Bar{\alpha_0}\Bar{\beta_1}\beta_0+\alpha_2\alpha_0\Bar{\beta_2}\beta_0+\alpha_2\alpha_0\beta_2\Bar{\beta_0}\label{eq6}
\end{split}
\end{equation}
\begin{equation}
\begin{split}
O_3=&\alpha_1\Bar{\alpha_0}\beta_2+\Bar{\alpha_2}\alpha_1\alpha_0\Bar{\beta_2}\beta_1\beta_0+\alpha_1\beta_2\Bar{\beta_1}\\
&+\alpha_2\Bar{\alpha_1}\beta_1+\alpha_2\alpha_0\beta_2\beta_0+\alpha_2\beta_1\Bar{\beta_0}\label{eq7}
\end{split}
\end{equation}
\begin{equation}
O_4=\alpha_1\alpha_0\beta_2\beta_1+\alpha_2\beta_2+\alpha_2\alpha_1\beta_1\beta_0\label{eq8}
\end{equation}
\begin{equation}
O_5=0\label{eq9}
\end{equation}

\begin{table}[htbp]
\caption{Approximate Truth Table for MUL$_{3\times3}$\_2 (Value\textgreater31)}
\begin{center}
\resizebox{\columnwidth}{!}{
\renewcommand{\arraystretch}{1.5}
\begin{tabular}{|c|c|c|c|c|c|c|c|c|c|c|}
\hline
\textbf{$\alpha_{2,1,0}$} & \textbf{$\beta_{2,1,0}$}&\textit{\textbf{Value}}&\textbf{$O_5$}&\textbf{$O_4$}&\textbf{$O_3$}&\textbf{$O_2$}&\textbf{$O_1$}&\textbf{$O_0$}&\textit{\textbf{Value'}}&\textit{\textbf{ED}}\\
\hline
101&111&35&0&1&1&0&1&1&27&8\\
\hline
\textbf{11}0&\textbf{11}0&36&\textbf{1}&\textbf{0}&1&0&0&0&\textbf{40}&\textbf{4}\\
\hline
\textbf{11}0&\textbf{11}1&42&\textbf{1}&\textbf{0}&1&1&1&0&\textbf{46}&\textbf{4}\\
\hline
111&101&35&0&1&1&0&1&1&27&8\\
\hline
\textbf{11}1&\textbf{11}0&42&\textbf{1}&\textbf{0}&1&1&1&0&\textbf{38}&\textbf{4}\\
\hline
\textbf{11}1&\textbf{11}1&49&\textbf{1}&\textbf{0}&1&1&0&1&\textbf{45}&\textbf{4}\\
\hline
\end{tabular}
}
\label{tab3}
\end{center}
\end{table}

Compared with the exact multiplier, the hardware cost of the above approximate multiplier, represented by \eqref{eq4}-\eqref{eq9}, is significantly lower. Detail comparison results will be shown in Section \ref{sec3}.
In order to reduce the \textbf{MED}, we revise the truth table of MUL$_{3\times3}$\_1 and adopt a prediction unit to determine values of $O_{5,4}$. We find that the common input bits corresponding to the four cases with larger \textbf{ED} are $\alpha_{2,1}=11$, $\beta_{2,1}=11$, as shown in TABLE \ref{tab3} in bold, which can be represented as $\alpha_2\cdot\alpha_1\cdot\beta_2\cdot\beta_1=1$. In those four cases, we set $O_5=1$, $O_4=0$ to construct MUL$_{3\times3}$\_2, reducing the \textbf{MED} from \textbf{1.125} to \textbf{0.5}. 

Although MUL$_{3\times3}$\_2 increases the number of outputs than MUL$_{3\times3}$\_1, it can reduce the power consumption with a small area overhead. The area, power consumption and delay results of MUL$_{3\times3}$\_1 and MUL$_{3\times3}$\_2 are given in Section \ref{sec3}.

\subsection{Aggregation for Large Multipliers}
\begin{figure}[htbp]
\centerline{\includegraphics[scale=0.23]{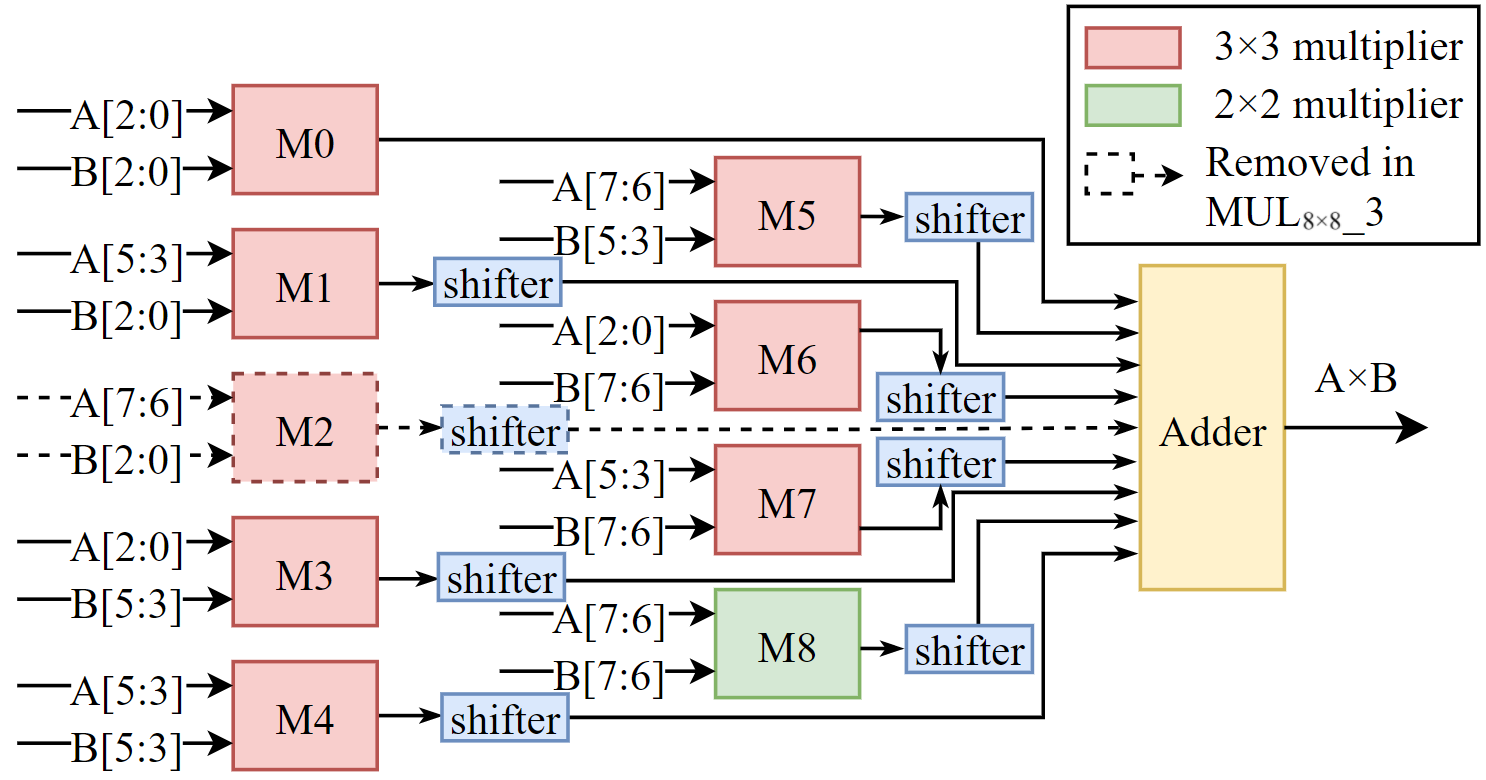}}
\caption{Block architecture of 8-bit multiplier based on 3×3 and 2×2 multipliers.}
\label{fig2}
\end{figure}

In order to customize the multiplier, we can learn that the input A or B tends to be small numbers after the LeNet quantization, where most of input values and weights of LeNet are in the range of (0,31) and (96,159), respectively. It is similar to other DNNs like AlexNet, ResNet-19, and VGG16. An 8×8 multiplier can be built by aggregating approximate 3×3 multipliers and 2×2 multipliers, as shown in Fig.\ref{fig2}, where A[7:0], B[7:0] are the inputs and $M_0$-$M_8$ are low bit-width multipliers to generate partial products.
Based on the above architecture, we can design three 8×8 approximate multipliers by selecting different 3-bit approximate multiplier designs as shown in TABLE ~\ref{tab5}. 

\begin{table}[htbp]
\caption{Aggregations of Three 8×8 Multipliers}
\begin{center}
\renewcommand{\arraystretch}{1.5}
\begin{tabular}{|c|c|c|}
\hline
\textit{\textbf{Name}} &\textit{ \textbf{M$_0$-M$_7$}}&\textit{\textbf{M$_8$}}\\
\hline
MUL$_{8\times8}$\_1&MUL$_{3\times3}$\_1&\multirow{3}{*}{Exact$_{2\times2}$} \\
\cline{1-2}
MUL$_{8\times8}$\_2&MUL$_{3\times3}$\_2&\\
\cline{1-2}
MUL$_{8\times8}$\_3$^{\mathrm{a}}$&MUL$_{3\times3}$\_2&\\
\hline
\multicolumn{3}{r}{$^{\mathrm{a}}$\textit{M$_2$} and the shifter are removed as shown in Fig.~\ref{fig2}.}
\end{tabular}
\label{tab5}
\end{center}
\end{table}


In addition, the DNN weights can be retrained to ensure that most weights are in (0, 31), where A[7:6] or B[7:6] is 00, so that we can remove M2 or M6 and the corresponding shifter from the architecture in Fig.\ref{fig2}, which can further reduce the critical path delay and power consumption.

\section{Experimental Results of Multipliers}\label{sec3}
\subsection{Error metrics}\label{AA}
In this paper, we use \textbf{ED}, \textbf{MED}, \textbf{ER}, \textbf{NMED}, and \textbf{MRED} as error metrics. \textbf{ED}, \textbf{
MED}, and \textbf{ER} are presented in \eqref{eq1}-\eqref{eq3}. \textbf{NMED} and \textbf{MRED} are defined as follows.

\begin{equation}
NMED=\frac{MED}{(2^n-1)^2}\label{eq10}
\end{equation}
\begin{equation}
MRED=\frac{ED}{Value'\cdot2^n}\label{eq11}
\end{equation}

TABLE \ref{tab6} shows the metric values of three approximate multipliers proposed in Section \ref{sec2} and the existing approximate multipliers. Average \textbf{ER}, \textbf{MED}, \textbf{NMED} and \textbf{MRED} of MUL$_{8\times8}$\_1,2,3 are 24.9\%, 298, 0.46\%, and 1.81\%, respectively. Such advantages of low \textbf{ER} and \textbf{NMED} can greatly improve the precision. 

\begin{table}[htbp]
\caption{Arithmetic Accuracy Comparision of Approximate Multipliers}
\begin{center}
\renewcommand{\arraystretch}{1.5}
\begin{tabular}{|c|c|c|c|c|}
\hline
\textit{\textbf{Name}} &\textit{ \textbf{ER(\%)}}&\textit{\textbf{MED}}&\textit{\textbf{NMED(\%)}}&\textit{\textbf{MRED(\%)}}\\
\hline
MUL$_{8\times8}$\_1 & 22.8 & 137.04 & 0.21 & 1.50\\
\hline
MUL$_{8\times8}$\_2 & 20.49 & 114.83 & 0.18 & 1.42\\
\hline
MUL$_{8\times8}$\_3 & 31.41 & 648.20 & 1.00 & 2.53\\
\hline
SiEi\cite{b7} & 31.59 & - & 0.20 & 0.62\\
\hline
PKM\cite{b10} & 49.86 & 938.32 & 1.44 & 3.89\\
\hline
ETM\cite{b12} & 98.88 & - & 2.85 & 25.21\\
\hline
SV\cite{b21} & - & - & 0.35 & 6.75\\
\hline
\end{tabular}
\label{tab6}
\end{center}
\end{table}

\begin{table}[htbp]
\caption{Comparision of 3×3 Approximate Multipliers in terms of Area, Power, Delay}
\begin{center}
\renewcommand{\arraystretch}{1.5}
\begin{tabular}{|c|c|c|c|}
\hline
\multirow{2}{*}{\textit{\textbf{Types}}} &\textit{ \textbf{Area( $\mu m^{2}$)}} & \textit{\textbf{Power(mW)}} &\textit{ \textbf{Delay(ns)}}\\ 
      & \textit{(Improvement)} & \textit{(Improvement)} &\textit{(Improvement)} \\ 
\hline
Exact (baseline)&67.68&	3.73&	0.45\\
\hline
MUL$_{3\times3}$\_1&	43.20 \textit{(36.17\%)}	&2.40 \textit{(35.66\%)}&	0.26 \textit{(42.22\%)}\\
\hline
MUL$_{3\times3}$\_2&46.44 \textit{(31.38\%)}&	2.36 \textit{(36.73\%)}&	0.26 \textit{(42.22\%)}\\
\hline
\end{tabular}
\label{tab4}
\end{center}
\end{table}

\begin{table}[htbp]
\caption{Comparision of 8×8 Approximate Multipliers in terms of Area, Power, Delay}
\begin{center}
\renewcommand{\arraystretch}{1.5}
\scalebox{0.97}{
\begin{tabular}{|c|c|c|c|}
\hline
\multirow{2}{*}{\textit{\textbf{Types}}} &\textit{ \textbf{Area( $\mu m^{2}$)}} & \textit{\textbf{Power(mW)}} &\textit{ \textbf{Delay(ns)}}\\ 
      & \textit{(Improvement)} & \textit{(Improvement)} &\textit{(Improvement)} \\ 
\hline
Exact (baseline)&744.59&	58.12 &	1.58\\
\hline
MUL$_{8\times8}$\_1&	596.16 \textit{(19.93\%)}& 45.66 \textit{(21.44\%)}	&1.29 \textit{(18.35\%)}\\
\hline
MUL$_{8\times8}$\_2& 646.92 \textit{(13.12\%)}&	50.84 \textit{(12.53\%)}&	1.41 \textit{(10.76\%)}\\
\hline
MUL$_{8\times8}$\_3& 571.32 \textit{(23.27\%)}&	42.28 \textit{(27.25\%)}&	1.29 \textit{(18.35\%)}\\
\hline
SiEi\cite{b7}& 579.51 \textit{(22.17\%)}&	39.57 \textit{(31.92\%)}&	1.37 \textit{(13.29\%)}\\
\hline
PKM\cite{b10}& 564.76 \textit{(24.15\%)}&	37.87 \textit{(34.84\%)}&	1.28 \textit{(18.99\%)}\\
\hline
\end{tabular}
}
\label{tab8}
\end{center}
\end{table}

\begin{table*}[htbp]
\caption{DNN Results of Approximate Multipliers}
\begin{center}
\renewcommand{\arraystretch}{1.5}
\begin{tabular}{|c|c|c|c|c|c|c|c|c|}
\hline
\textbf{\textit{Dataset}}&\multicolumn{3}{c|}{MNIST}&\multicolumn{5}{c|}{CIFAR-10}\\
\hline
\textit{\textbf{Multiplier Type}} &\textit{ \textbf{LeNet}}&\textit{\textbf{Regularization}} & \textbf{\textit{LeNet+}}&\textit{ \textbf{LeNet}}& \textbf{\textit{LeNet+}}&\textit{\textbf{VGG16}} &\textit{\textbf{AlexNet}}&\textit{\textbf{ResNet-19}}\\
\hline
Exact(baseline)&99.32\%&99.41\%&99.51\%&75.09\%&76.80\%&93.51\%&88.48\%&88.92\%\\
\hline
MUL$_{8\times8}$\_1&98.98\%&99.02\%&99.14\% &60.84\%&61.47\%&29.59\% &56.95\%&39.84\%\\
\hline
MUL$_{8\times8}$\_2&\textbf{99.32\%}&\textbf{99.41\%}&\textbf{99.51\%} &\textbf{74.21\%}&\textbf{77.00\%}&\textbf{91.60\%}&\textbf{88.17\%}&\textbf{87.58\%} \\
\hline
MUL$_{8\times8}$\_3&98.49\%&99.12\%&99.22\% &54.70\%&64.44\%&31.32\% &73.80\%&73.76\%\\
\hline
SiEi\cite{b7}&74.88\%&95.46\%&95.77\%&12.56\%&13.56\%&9.80\%&8.64\%&10.59\%\\
\hline
PKM\cite{b10}&96.32\%&98.39\%&98.40\%&36.23\%&47.05\%&11.47\%&37.95\%&24.72\%\\
\hline
\end{tabular}
\label{tab9}
\end{center}
\end{table*}

Because the \textbf{ER} and \textbf{MRED} of ETM \cite{b12} are too poor to compare and SV\cite{b21} lacks \textbf{ER} and \textbf{MED}, we select the five approximate multipliers: MUL$_{8\times8}$\_1-3, SiEi\cite{b7} and PKM\cite{b10} to carry out the follow-up experimental evaluation.

\subsection{Results of 3$\times$3 Approximate Multipliers}
In this part, all approximate multiplier architectures are written in Verilog, and synthesized by Synopsys Design Compiler (DC) with ASAP 7nm standard cell library\cite{b22} to obtain the area and power values.  
The DC results of 3×3 approximate multipliers are shown in TABLE VI. In comparison to the exact multiplier produced by Synopsys DesignWare library, MUL$_{3\times3}$\_1 and MUL$_{3\times3}$\_2 can reduce the area by 36.17\% and 31.38\% and power consumption by 35.66\% and 36.73\%, respectively. The delay can be reduced by 42.22\%.

\subsection{Results of 8$\times$8 Approximate Multipliers}
We can obtain DC results of 8×8 approximate multiplier as shown in TABLE \ref{tab8}, where MUL$_{8\times8}$\_1,2,3 can reduce the area by 19.93\%, 13.12\%, and 23.27\%, respectively. The power consumption is reduced by 21.44\%, 12.53\%, and 27.25\%, respectively. On the critical delay, the proposed designs can also reduce 18.35\%, 10.76\%, and 18.35\%.
In \cite{b10}, PKM uses the approximate unit in MSBs, which can reduce the area, power, and delay to some extent. However, it leads to the high \textbf{DAL} as shown in the following Section \ref{sec4}.

\section{DNN with Approximate Multipliers}\label{sec4}
In this section, the DNN platform\cite{b17} is extended to evaluate the \textbf{DAL} caused by replacing exact multipliers with approximate 8×8 multipliers. The 8-bit unsigned approximate multipliers in Section \ref{sec2} are evaluated by this platform and compared with the exact multiplier. When the DNN accuracy is lower than 40\%, we usually think that the network with approximate multipliers loses its predictive ability.

Under this platform, we train the DNN on the MNIST and CIFAR10 datasets, respectively. Based on the DNN structure, we can increase the number of convolution layers in LeNet to increase network complexity, which can also improve the recognition ability of the input layer. The modified LeNet structure can be denoted by \textbf{LeNet+}. And the tolerance of approximate multipliers in larger networks can be predicted.  

On the MNIST, MUL$_{8\times8}$\_1,2,3 are good-performance approximation multipliers. Under the original LeNet structure, the worst \textbf{DAL} of our three designs is  MUL$_{8\times8}$\_3 (0.83\%), and the best one is MUL$_{8\times8}$\_2 (no accuracy loss).

We find that the retraining we applied can reduce the \textbf{DAL} of MUL$_{8\times8}$\_1,2,3 when compared to the baseline in TABLE \ref{tab9}. Regularization and \textbf{LeNet+} keep the DNN accuracy of MUL$_{8\times8}$\_2  at a high level and they can improve the \textbf{DAL} of the worst design to 0.39\% and 0.37\%, respectively. It also follows that \textbf{LeNet+} has a better improvement than regularization. So we apply \textbf{LeNet+} only on CIFAR10.

The original DNN accuracy of MUL$_{8\times8}$\_2 is similar to that of baseline on CIFAR10, and in \textbf{LeNet+}, it even exceeds the DNN accuracy of the exact multiplier. However, PKM, which has a moderate effect on MNIST, has only 36.23\% DNN result on CIFAR10, which can reach 47.05\% after retraining in \textbf{LeNet+}. 

LeNet is sufficient for MNIST, but for CIFAR10, we need to use larger DNNs to evaluate the approximate multipliers. VGG16, AlexNet, and ResNet-19 are then adopted to our DNN platform. MUL$_{8\times8}$\_2 can obtain better \textbf{DAL} than other designs in our evaluation. It can be seen that the prediction unit in MUL$_{3\times3}$\_2 is very necessary, and it consumes a little extra area to achieve the improvement of accuracy. In addition to reproduced multipliers above, we choose another state-of-the-art multiplier to compare. The \textbf{DAL} of the multiplier design shown in\cite{b23} is 3.86\% in VGG16 ($m=3$) on CIFAR10, while MUL$_{8\times8}$\_2 can obtain a lower \textbf{DAL} (1.91\%).

\section{Conclusion}\label{sec5}
In this paper, we propose two 3×3 approximate multipliers, which have the same \textbf{ER} but different \textbf{MEDs}. Compared with the exact 3×3 multiplier, MUL$_{3\times3}$\_1,2 can reduce the area, power consumption, and delay by 33.5\%, 36.1\% and 42.2\% on average, respectively. They are then aggregated with 2×2 multipliers into the 8×8 multiplier according to DNN weights, which is evaluated and co-optimized by the extended DNN platform. MUL$_{8\times8}$\_2 has no \textbf{DAL} for LeNet on MNIST dataset without retraining. On CIFAR10, MUL$_{8\times8}$\_2 even exceeds the DNN accuracy of exact multiplier on \textbf{LeNet+}. For ResNet-19, AlexNet, and VGG16, the \textbf{DALs} of MUL$_{8\times8}$\_2 are significantly smaller than those of other designs on CIFAR10. Overall, due to the hardware and software co-optimization, the \textbf{DAL} can be negligible with the proposed MUL$_{8\times8}$\_2.


\end{document}